\documentclass{elsart5p}
\usepackage{graphics}
\usepackage{graphicx}
\usepackage{epsfig}
\usepackage{amssymb}
\usepackage{amsmath}
\usepackage{color}
\usepackage{lineno}

\newcommand{\Szero}{\mbox{$^{1\!}S_0$}}

\newcommand{\fmn}[2]{\mbox{${\textstyle \frac{#1}{#2}}$}}

\begin{document}

\begin{frontmatter}

\title{Study of the $p\pol{d}\to n\{pp\}_{s}$ charge-exchange reaction using a polarised deuterium target}

\author[imp,snst,ucas]{B.~Gou},
\author[hepi,ikp]{D.~Mchedlishvili\corauthref{cor1}}, \ead{d.mchedlishvili@fz-juelich.de}\corauth[cor1]{Corresponding author.}
\author[hepi,ikp]{Z.~Bagdasarian},
\author[pnpi]{S.~Barsov},
\author[hepi,ikp]{D.~Chiladze},
\author[jinr1,ikp]{S.~Dymov},
\author[ikp]{R.~Engels},
\author[ikp]{M.~Gaisser},
\author[ikp]{R.~Gebel},
\author[ikp,pnpi]{K.~Grigoryev},
\author[ikp]{M.~Hartmann},
\author[ikp]{A.~Kacharava},
\author[munster]{A.~Khoukaz},
\author[krakow]{P.~Kulessa},
\author[jinr1]{A.~Kulikov},
\author[ikp]{A.~Lehrach},
\author[imp]{Z.~Li},
\author[hepi]{N.~Lomidze},
\author[ikp]{B.~Lorentz},
\author[hepi,jinr1]{G.~Macharashvili},
\author[ikp,jinr1]{S.~Merzliakov},
\author[munster]{M.~Mielke},
\author[ikp,pnpi]{M.~Mikirtychyants},
\author[ikp,pnpi]{S.~Mikirtychyants},
\author[hepi]{M.~Nioradze},
\author[ikp]{H.~Ohm},
\author[ikp]{D.~Prasuhn},
\author[ikp]{F.~Rathmann},
\author[ikp]{V.~Serdyuk},
\author[ikp]{H.~Seyfarth},
\author[jinr1]{V.~Shmakova},
\author[ikp]{H.~Str\"oher},
\author[hepi]{M.~Tabidze},
\author[ikpros,skob]{S.~Trusov},
\author[jinr1]{D.~Tsirkov},
\author[jinr1,msu]{Yu.~Uzikov},
\author[ikp,pnpi]{Yu.~Valdau},
\author[snst]{T.~Wang},
\author[ferr]{C.~Weidemann},
\author[ucl]{C.~Wilkin},
\author[imp]{X.~Yuan}

\address[imp]{Institute of Modern Physics, Chinese Academy of Sciences, Lanzhou 730000, China}
\address[snst]{School of Nuclear Science and Technology, Lanzhou University, Lanzhou 730000, China}
\address[ucas]{University of Chinese Academy of Sciences, Beijing 100049, China}
\address[ikp]{Institut f\"ur Kernphysik and J\"ulich Centre for Hadron Physics, Forschungszentrum J\"ulich, D-52425 J\"ulich, Germany}
\address[hepi]{High Energy Physics Institute, Tbilisi State University, GE-0186 Tbilisi, Georgia}
\address[pnpi]{High Energy Physics Department, Petersburg Nuclear Physics Institute, RU-188350 Gatchina, Russia}
\address[jinr1]{Laboratory of Nuclear Problems, JINR, RU-141980 Dubna, Russia}
\address[munster]{Institut f\"ur Kernphysik, Universit\"at M\"unster, D-48149 M\"unster, Germany}
\address[krakow]{H.~Niewodnicza\'{n}ski Institute of Nuclear Physics PAN, PL-31342 Krak\'{o}w, Poland}
\address[ikpros]{Institut f\"ur Kern- und Hadronenphysik, Forschungszentrum Rossendorf, D-01314 Dresden, Germany}
\address[skob]{Skobeltsyn Institute of Nuclear Physics, Lomonosov Moscow State University, RU-119991 Moscow, Russia}
\address[msu]{Department of Physics, M.~V.~Lomonosov Moscow State University, RU-119991 Moscow, Russia}
\address[ferr]{University of Ferrara and INFN, I-44100 Ferrara, Italy}
\address[ucl]{Physics and Astronomy Department, UCL, Gower Street, London, WC1E 6BT, UK}

\date{Received: \today / Revised version:}

\begin{abstract}
The vector and tensor analysing powers, $A_y$ and $A_{yy}$, of the
$p\pol{d}\to n\{pp\}_{s}$ charge-exchange reaction have been measured at a
beam energy of 600~MeV at the COSY-ANKE facility by using an unpolarised
proton beam incident on an internal storage cell target filled with polarised
deuterium gas. The low energy recoiling protons were measured in a pair of
silicon tracking telescopes placed on either side of the target. Putting a
cut of 3~MeV on the diproton excitation energy ensured that the two protons
were dominantly in the \Szero\ state, here denoted by $\{pp\}_{s}$. The
polarisation of the deuterium gas was established through measurements in
parallel of proton-deuteron elastic scattering. By analysing events where
both protons entered the same telescope, the charge-exchange reaction was
measured for momentum transfers $q\geq 160$~MeV/$c$. These data provide a
good continuation of the earlier results at $q\leq 140$~MeV/$c$ obtained with
a polarised deuteron beam. They are also consistent with impulse
approximation predictions with little sign evident for any modifications due
to multiple scatterings.
\end{abstract}


\end{frontmatter}

It was pointed out several years ago that the charge exchange of polarised
deuterons on hydrogen, $\pol{d}p\to \{pp\}_{\!s}n$, can furnish useful
information on the spin dependence of elastic neutron-proton amplitudes near
the backward centre-of-mass direction provided that the final proton pair
$\{pp\}_{\!s}$ is detected at very low excitation energy
$E_{pp}$~\cite{BUG1987}. In this limit the diproton is dominantly in the
\Szero\ state and so there is then a spin-isospin flip from the $(S,T)=(1,0)$
of the deuteron to the $(0,1)$ of the diproton. At small momentum transfers
between the deuteron and diproton, the deuteron charge-exchange amplitudes
can be interpreted in impulse approximation in terms of $np$ amplitudes times
form factors that reflect the overlap of the deuteron bound-state and the
diproton scattering-state wave functions~\cite{BUG1987}.

Following pioneering experiments at Saclay~\cite{ELL1987,KOX1993}, the most
detailed studies of the $\pol{d}p\to \{pp\}_{\!s}n$ reaction were undertaken
by the ANKE collaboration at deuteron energies of $T_d=1.2$, 1.6, 1.8, and
2.27~GeV, i.e., at energies per nucleon of $T_N=600$, 800, 900, and
1135~MeV~\cite{CHI2006,MCH2013}. At the three lower energies the
predictions~\cite{CAR1991} of the impulse approximation model describe the
data very well on the basis of $np$ input taken from the SAID SP07 partial
wave solution~\cite{ARN2007}. Deviations were, however, noted in the 2.27~GeV
data~\cite{MCH2013} that were ascribed to an overestimate of the strength of
the $np$ spin-longitudinal amplitude at 1135~MeV.

The major constraint on the ANKE programme is the maximum deuteron energy of
2.3~GeV available at the COSY accelerator~\cite{KAC2005}. To continue the
studies at COSY to higher energies, where there is great uncertainty in the
neutron-proton amplitudes, the experiments have to be carried out in inverse
kinematics, with a proton beam incident on a polarised deuterium target. The
study of the charge exchange at low momentum transfers would then require the
measurement of two low energy protons recoiling from the
target~\cite{MCH2013a}. We here report on the first measurement of the
$p\pol{d}\to n\{pp\}_{s}$ charge-exchange at 600~MeV that extends the earlier
deuteron beam data out to larger values of the momentum transfer $q$. Such a
programme clearly first necessitates a reliable determination of the
polarisation of the deuterium target and this will therefore be an important
element of this letter.

The experiment was carried out using the ANKE magnetic
spectrometer~\cite{BAR2001} situated inside the storage ring of the COoler
SYnchrotron (COSY)~\cite{MAI1997} of the Forschungszentrum J\"ulich. The
whole target facility consists of three major components: the atomic beam
source (ABS)~\cite{MIK2013}, the storage cell (SC)~\cite{GRI2007,GRI2014},
and the Lamb-shift polarimeter (LSP)~\cite{ENG2003}.

\begin{figure}[h]
\includegraphics[width=0.65\linewidth]{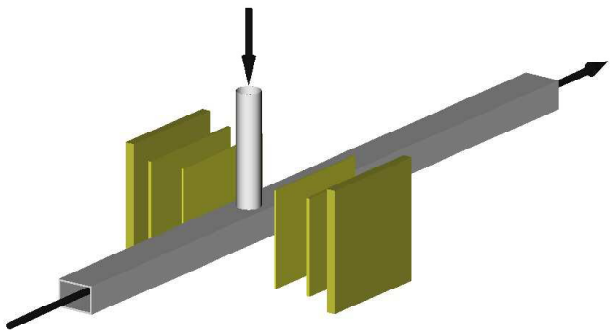}
\vspace{-3.5cm}
\caption{Schematic view of the ANKE target area showing the positions of the
polarised deuterium cell target and its feeding tube and the two Silicon 
Tracking Telescopes (STT). The beam direction is indicated by
the long horizontal arrow.} \label{targetarea}
\end{figure}

The ABS is capable of providing deuterium beams with different combinations
of vector ($Q_y$) and tensor ($Q_{yy}$) polarisations. Four different modes
were used in the current experiment with ideal polarisations of $(Q_y,Q_{yy})
= (+1, +1)$, $(-1, +1)$, $(0, -2)$ and $(0, +1)$, where the quantisation axis
$y$ is taken to be the upward normal to the plane of the COSY accelerator.
The atomic beams from the ABS are led into the storage cell placed inside the
ANKE vacuum chamber via a feeding tube and diffuse along the cell that is
illustrated in Fig.~\ref{targetarea}. The Lamb-shift polarimeter, which
measures the polarisation of the atomic beam from the ABS, is used for tuning
the settings of the ABS before the experiments.

The polarised deuterium gas cell was rather similar to that used in the
previous ANKE experiment with polarised hydrogen~\cite{MCH2013}. The cell was
made of $25~\mu$m thick aluminium foil (99.95\% Al) with the inner walls
coated with Teflon in order to minimise the depolarisation of the deuterium
atoms. The cell had dimensions
$20\times15\times390~$mm$^3$~\cite{GRI2007,GRI2014}. Such a cell increases
the target thickness by about two orders of magnitude compared to using the
ABS jet directly as a target and, as a result, an average luminosity of $L
\approx 5 - 7 \times 10^{28}$~cm$^{-2}$~s$^{-1}$ was obtained over the ten
days of data taking.

Though several nuclear reactions can be measured by detecting fast particles
that pass through the ANKE magnetic analyser, both the polarimetry and the
measurement of the charge exchange reaction were achieved by detecting only
slow particles that emerge from the target cell using a pair of silicon
tracking telescopes (STT)~\cite{SCH2003}.

The two STT, each consisting of three double-sided silicon strip layers of
70~$\mu$m, 300~$\mu$m and 5~mm thickness, were placed symmetrically inside
the vacuum chamber, to the left and right of the cell, as shown in
Fig.~\ref{targetarea}. The distances of the sensitive layers away from the
target axis were 2.8~cm, 4.8~cm, and 6.1~cm so that the STT covered the
laboratory polar angles $75^{\circ} < \theta_{\rm lab} < 140^{\circ}$. In
order to pass through the three layers, the recoiling protons or deuterons
must have energies of at least 2.5~MeV, 6~MeV, and 30~MeV, respectively. For
stopping particles the particle identification is unambiguous. In the case of
proton-deuteron elastic scattering, which is the main polarimetry reaction
used for this study, greater precision in the angle of the recoiling deuteron
is achieved by deducing it from the energy measured in the telescope rather
than from a direct angular measurement.

The experiment was conducted using the pairs of polarisation modes that are
defined in Table~\ref{polar}. The ABS was configured in such a way as to
provide identical polarised gas densities for the pairs of modes (1,2) and
(3,4). The target was switched every 10 seconds, first between polarisation
modes 1 and 2 and later between 3 and 4. Since the beam was stable on such a
short time scale, this procedure ensured equal luminosities for each member
of the pair.

\begin{table}[h!]
\centering
\begin{tabular}{|c|c|c|c|c|c|c|}
\hline
 Pol. & \multicolumn{3}{c|}{Modes 1,2} & \multicolumn{3}{c|}{Modes 3,4} \\
\cline{2-7}
     & Ideal & Measured & Sys. err. & Ideal & Measured & Sys. err.\\
\hline
 $\Delta Q_y$ & $+2$ & $\phantom{-}1.46\pm0.01$ & $0.03$ & $\phantom{-}0$ & $-0.07\pm0.01$ & $0.01$\\
 $\langle Q_{y}\rangle$ & $\phantom{-}0$ & $-0.03\pm0.01$ & $0.01$ & $\phantom{-}0$ & $-0.02\pm0.02$ & $0.01$\\
 $\Delta Q_{yy}$ & $\phantom{-}0$ & $\phantom{-}0.17\pm0.02$ & $0.01$ & $-3$ & $-1.68\pm0.02$ & $0.14$\\
 $\langle Q_{yy}\rangle$ & $+1$ & $\phantom{-}0.88\pm0.03$ & $0.11$ & $-\fmn{1}{2}$ & $-0.13\pm0.06$ & $0.03$\\
\hline
\end{tabular}
\vspace{5mm} \caption{The ideal and measured polarisations of the target
given in terms of average polarisations, $\langle Q\rangle$, and the
polarisation differences, $\Delta Q$, between the members of the two pairs of
polarised modes used in the ANKE experiment. The systematic uncertainties,
arising from the analysing powers of the proton-deuteron elastic reaction,
are listed separately. It is important to note that the effective target
thicknesses are identical in the (1,2) modes, as they are also in the (3,4)
modes. \label{polar}}
\end{table}

\begin{figure}[h]
\begin{center}
\includegraphics[width=0.45\textwidth]{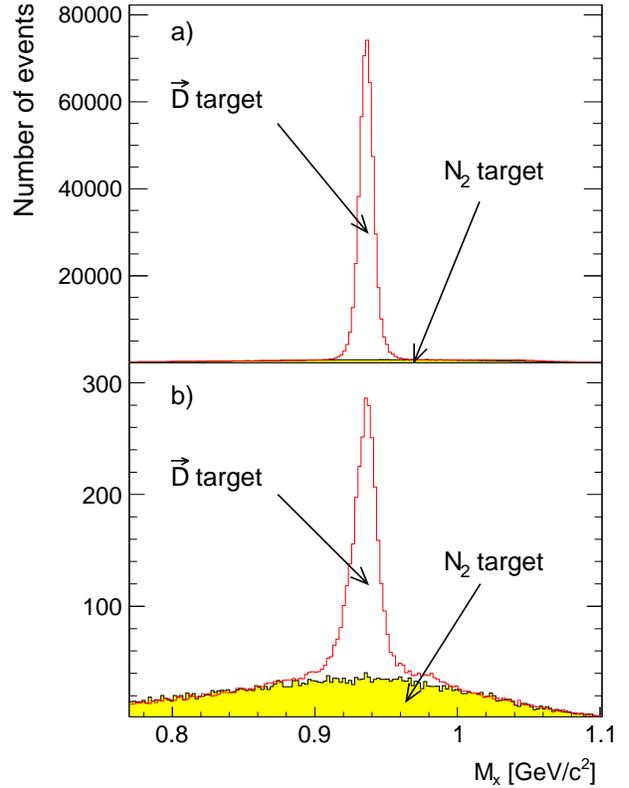}
\caption{The missing-mass $M_X$ spectra of the $p\pol{d}\to
dX$ (a) and $p\pol{d}\to ppX$ (b) reaction measured in the STT for 600~MeV
protons incident on the polarised cell target. In both cases the shape of the
background was simulated by filling the cell with unpolarised nitrogen gas.}
\label{mmsq}
\end{center}
\end{figure}

Due to the loss of polarisation of the atoms in the target through collisions
with the cell walls or through the recombination into molecules, the
polarisation of the deuterium in the cell is smaller than that of the atoms
coming from the ABS. The values of the target polarisations have therefore to
be established under the actual conditions of the experiment and, for this
purpose, elastic proton-deuteron scattering was measured, with the recoiling
deuteron being detected in the STT. It is here very important to be sure that
the reaction had taken place on the target gas rather than on the aluminium
walls. To provide a rapid simulation of this background, the cell was later
filled with unpolarised nitrogen gas. As shown in Fig.~\ref{mmsq}a, this
gives a very good description of the background away from the missing-mass
peaks that are associated with the unobserved proton coming from the
deuterium target. In the $pd$ elastic scattering case the background is in
any case very low; there is no difficulty at all in identifying elastic
events because of the strong link between the angle and the energy deposited
in the STT.

Elastic proton-deuteron scattering has a high cross section at small angles
and, though this varies fast with momentum transfer, this quantity is very
well determined in the STT. There are measurements of both the vector $A_y$
and tensor $A_{xx}$ and $A_{yy}$ analysing powers of the $\pol{d}p\to dp$
reaction with polarised deuteron beams at neighbouring energies to our
600~MeV per nucleon. The data from Argonne at $T_d=1194$~MeV~\cite{HAJ1987},
from SATURNE at $T_d=1198$~MeV~\cite{ARV1984,ARV1988}, and from ANKE at
$T_d=1170$~MeV~\cite{CHI2006a} show strong and well measured analysing
powers.

For a two-body reaction, such as $p\pol{d}$ elastic scattering, or a more
general process, such as $p\pol{d}\to n\{pp\}_{s}$, where one does not
consider the internal variables of the diproton, the number of particles $N$
scattered at polar angle $\theta$ and azimuthal angle $\phi$ is given by
\begin{eqnarray}\label{general}\nonumber
N(\theta,\phi) &=& N^0(\theta)\left\{1+\fmn{3}{2}Q_yA_y^d(\theta)\cos\phi\right.\\
&&\hspace{-1.2cm}\left.+\fmn{1}{4}Q_{yy}[A_{yy}(\theta)(1+\cos 2\phi
)+A_{xx}(\theta)(1-\cos 2\phi )]\right\}\!,
\end{eqnarray}
where $\phi$ is measured from the horizontal plane of the COSY
accelerator. Here $N^0$ is the corresponding number obtained
with an unpolarised beam.

The STT~\cite{SCH2003}, which have limited angular acceptance, are placed in
the same horizontal plane as the target cell and, under these conditions,
only accept events close to $\phi =0^{\circ}$ and $180^{\circ}$. As a
consequence, the present measurements are primarily sensitive to the values
of $A_y$ and $A_{yy}$ for any reaction. During the experiment the working
conditions for the STT were such that the difference between the total
efficiencies for different polarised modes in a pair is expected to be very
small. This was experimentally verified using background events that were
free from polarisation effects. Such events were collected from the vicinity
of the missing-mass peak corresponding to the elastic $pd$ scattering of
Fig.~\ref{mmsq}a. By building the count ratios between the two members of a
polarisation pair, which is directly the product of relative efficiency times
the relative luminosity between the two modes, it was shown that the relative
efficiency is unity within 1.5\%. We note here again that the luminosities
are the same for each of the polarisation modes in a given pair of
Table~\ref{polar}.

Using the data from the left and right STT separately, the ratios of the
difference to the sum of counts were built for each pair of polarised modes.
We describe here the procedure used for modes (1,2); modes (3,4) were treated
in a similar fashion. The (1,2) ratios correspond to:
\begin{equation}
\frac{N_1-N_2}{N_1+N_2} = \frac{\Delta V + \Delta T}
{2(1 + \langle V\rangle + \langle T\rangle)}\,,
\label{polratio}
\end{equation}
where $N_1$ and $N_2$ are the number of counts in modes 1 and 2. In terms of
polarisation observables,
\begin{eqnarray}
\centering
\nonumber
\Delta V &=& \fmn{3}{2}\Delta Q_yA_{y}^d(\theta)\cos\phi \,,\\
\nonumber
\langle V\rangle &=& \fmn{3}{2}\langle Q_y\rangle A_{y}^d(\theta)\cos\phi \,,\\
\nonumber
\Delta T &=& \fmn{1}{4}\Delta Q_{yy}\left[A_{yy}(\theta)(1+\cos2\phi)
+ A_{xx}(\theta)(1-\cos2\phi)\right]\!,\\
\nonumber
\langle T\rangle &=& \fmn{1}{4}\langle Q_{yy}\rangle\left[A_{yy}(\theta)(1+\cos2\phi)
+ A_{xx}(\theta)(1-\cos2\phi)\right]\!,\\
\label{polvar}
\end{eqnarray}
where $\Delta Q = Q_{1}-Q_{2}$ and $\langle Q\rangle = (Q_{1}+Q_{2})/2$ are,
respectively, the difference and the average polarisations for the (1,2)
pair. The two-dimensional $(\theta,\phi)$ maps were built from these ratios
for $pd$ elastic scattering, which were then fitted simultaneously in both
variables in order to determine all four polarisation values entering in
Eq.~(\ref{polvar}).

The vector ($A_y$) and tensor ($A_{xx}$, $A_{yy}$) analysing powers used in
the polarimetry were taken as weighted averages of the measurements already
mentioned~\cite{HAJ1987,ARV1984,ARV1988,CHI2006a} that were carried out at
very close energies per nucleon. Since Eq.~(\ref{general}) shows that the
polarised cross section is significantly less sensitive to the tensor than
the vector term, $Q_{yy}$ is less well determined than $Q_y$. The results
listed in Table~\ref{polar} clearly illustrate this behaviour. Moreover, the
systematic uncertainties, arising from the determination of the input
analysing powers, are significantly larger for the tensor polarisations.
Estimates for these uncertainties are also given in Table~\ref{polar}.

\begin{figure}[h]
\begin{center}
\includegraphics[width=0.45\textwidth]{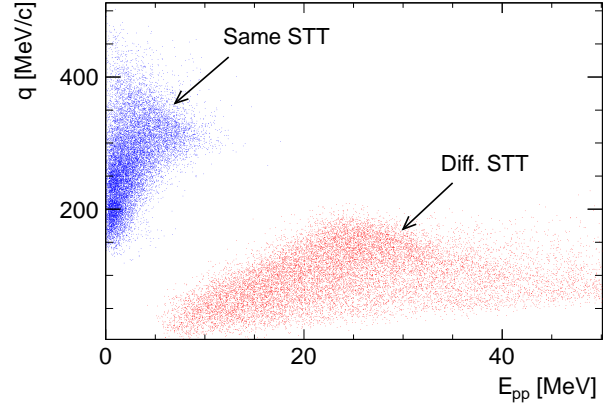}
\caption{The three-momentum transfer $q$ versus the $pp$ excitation energy
$E_{pp}$ for the $p\pol{d}\to \{pp\}_{\!s}X$ events at $T_p=600$~MeV that
fall within $\pm3\sigma$ of the neutron peak. The data are shown separately
for cases where the two protons enter the same (blue) or different (red) STT.
The current construction of the STT means that there can be no events where
$q$ and $E_{pp}$ are simultaneously small.} \label{eppvsq}
\end{center}
\end{figure}

When measuring the $\pol{d}p \to \{pp\}_{\!s}n$ reaction with a polarised
deuteron beam by detecting the two fast protons in the ANKE forward detector,
it was possible to investigate regions where the momentum transfer $q$
between the deuteron and the diproton and the diproton energy $E_{pp}$ were
both small~\cite{MCH2013}. This is no longer the case in inverse kinematics
when the two slow protons are measured in the STT. This is due to the
requirement that the protons pass through the first silicon layer of the
detector. They then have energies above 2.5~MeV, i.e., momenta above about
70~MeV/$c$. This means that, if the two protons are measured in different
STT, then $q$ can be small but $E_{pp}\gtrsim 6$~MeV because the protons are
going in opposite directions. On the other hand, if the two protons are
measured in the same STT, then $E_{pp}$ can be small but the momentum
transfer has a lower limit of $q\gtrsim 2\times 70$~MeV/$c$. There is
therefore a significant hole in the acceptance, which is demonstrated by the
data shown in Fig.~\ref{eppvsq}. This is in complete contrast to the deuteron
beam data, where the region where $E_{pp}<3$~MeV and $0<q< 140$~MeV/$c$ is
routinely accessed~\cite{MCH2013}.

In this letter we report only on results obtained at low $E_{pp}$ for events
where the two protons entered the same STT. These data provide a natural
continuation from the small $q$ region studied in the deuteron beam
measurements~\cite{MCH2013}. Having detected two protons in one STT the
missing mass of the $p\pol{d}\to ppX$ reaction is constructed and an example
of this is shown in Fig.~\ref{mmsq}b. For this three-body final state the
measurement errors are larger than for $pd$ elastic scattering and the
background from the cell walls can be more problematic. However, the shape of
this background is simulated very well by the contribution from the nitrogen
gas filling that is also shown. Similar background subtractions were made for
all four polarisation modes of Table~\ref{polar}, where the nitrogen
normalisation was fixed from fitting outside the peak region.

\begin{figure}[h]
\begin{center}
\includegraphics[width=0.45\textwidth]{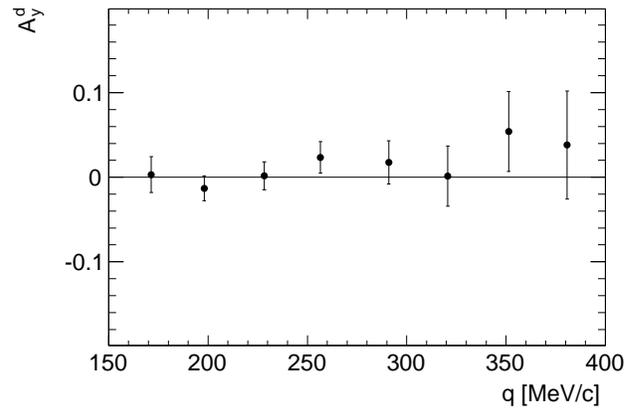}
\caption{Deuteron vector analysing power $A_{y}^{d}$ of the $p\pol{d}\to
n\{pp\}_{s}$ reaction with an $E_{pp}< 3$~MeV cut. Impulse approximation
predictions~\cite{CAR1991} based upon the SP07 solution for the
neutron-proton elastic scattering amplitudes~\cite{ARN2007} were used to
correct for tensor analysing power effects.} \label{Ayd}
\end{center}
\end{figure}

Data were taken by flipping the target polarisations between modes (1,2) or
between (3,4). Since $\Delta Q_y$ is largest for the (1,2) pair, this
combination provides the best measurement of the deuteron vector analysing
power $A_y^d$ in the charge exchange reaction, using the same technique as
that employed for the polarimetry in $pd$ elastic scattering. Corrections for
the tensor analysing powers effects in the charge-exchange reaction were made
using impulse approximation predictions~\cite{CAR1991}. As shown in
Fig.~\ref{Ayd}, for the standard cut of $E_{pp}<3$~MeV it was found that
$A_y^d=0$ within error bars, with an average over all momentum transfers of
$\langle A_y^d\rangle =0.005\pm 0.008$. This agrees with theoretical
predictions~\cite{BUG1987} and experimental results at lower momentum
transfers~\cite{CHI2006,MCH2013}.

\begin{figure}[h]
\begin{center}
\includegraphics[width=0.45\textwidth]{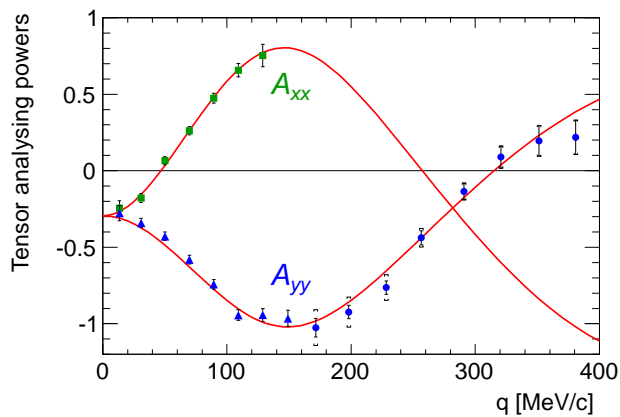}
\caption{Tensor analysing powers $A_{xx}$ (green squares) and $A_{yy}$ (blue
triangles) of the $\pol{d}p\to \{pp\}_{s}n$ reaction with $E_{pp}< 3$~MeV
from the earlier deuteron beam measurements at
600~MeV/nucleon~\cite{CHI2006}. These data were restricted to the region
$q<160$~MeV/$c$. The current $A_{yy}$ results (blue dots), which provide
data in the region $q > 160$~MeV/$c$, were obtained in inverse kinematics
using a 600~MeV unpolarised proton beam incident on a polarised deuterium
target. The extended error bars indicated by $\sqcup$ and $\sqcap$, which are
important in the STT data below 250~MeV/$c$, include the effects arising from
the uncertainties in the target polarisations.
The curves are impulse approximation predictions~\cite{CAR1991} based upon
the SP07 solution for the neutron-proton elastic scattering
amplitudes~\cite{ARN2007}.} \label{Ayyppn}
\end{center}
\end{figure}

The best determination of the tensor analysing power $A_{yy}$ is found by
comparing the rates in modes (3,4), which have the same luminosities but
tensor polarisations of opposite sign. The results with the standard $E_{pp}$
cut are shown in Fig.~\ref{Ayyppn}, where they are compared with ANKE data at
lower momentum transfers~\cite{CHI2006} and also with impulse approximation
predictions~\cite{CAR1991}. With the $E_{pp}<3$~MeV cut it is believed that
the \Szero\ state dominates at small $q$ but $P$ and higher waves become more
important as $q$ increases. A tighter cut on $E_{pp}$ would, in principle, be
possible since the $E_{pp}$ resolution is around $0.3$~MeV for $E_{pp}<1$~MeV
and below $1$~MeV for higher $E_{pp}$. However the count rate drops rapidly
with a lower $E_{pp}$ cut and the currently available data might not help in
the identification of any possible dilution of $A_{yy}$ by the higher partial
waves.

The new $A_{yy}$ results shown in Fig.~\ref{Ayyppn} are very large below
about 200~MeV/$c$ but join quite smoothly onto the lower momentum transfer
data obtained with the polarised deuteron beam~\cite{CHI2006}. Furthermore
the data seem to be essentially consistent with impulse approximation
predictions~\cite{CAR1991}. However, at such large values of $q$ one has to
question to what extent the single scattering of impulse approximation is
still quantitatively valid. Formulae have been derived that incorporate the
effects of double scattering but only for the \Szero\ final
state~\cite{BUG1987}. Such terms have little effect for momentum transfers
below 140~MeV/$c$ where $A_{yy}$ has its minimum but double scattering in the
\Szero\ limit tends to push the momentum transfer for which $A_{yy}$ crosses
zero down by about 20~MeV/$c$. On the other hand, double scattering will be
far less important for $P$ and higher waves in the $pp$ system and so the
20~MeV/$c$ shift must be considered as very much an upper limit. More
detailed calculations are in progress.

Data taken with two separate STT have necessarily large $E_{pp}$, which will
generally reduce the analysing power signal through the excitation of higher
partial waves~\cite{CAR1991}. Unfortunately, the data so far obtained have
very limited statistics and could not usefully determine the tensor analysing
power at high $E_{pp}$.

Measurements at higher energies are scheduled for the near future and these
will also include studies with polarised proton beams in order to determine
spin-correlation coefficients. After the installation of a Siberian snake at
COSY, it may even be possible to study spin-longitudinal--spin-transverse
correlations. Another attractive possibility is to measure in coincidence
fast protons or pions in the ANKE magnetic spectrometer. These would allow
one to study in detail the $\{pp\}_{\!s}\Delta^0(1232)$ final state, where
the decay $\Delta^0(1232)\to \pi^-p$ defines the
alignment of the isobar~\cite{KAC2005}.\\[1ex]

We are grateful to other members of the ANKE Collaboration for their help
with this experiment and to the COSY crew for providing such good working
conditions, especially in respect of the polarised deuterium cell. The values
of the SAID neutron-proton amplitudes were kindly furnished by
I.I.~Strakovsky. This work has been partially supported by the
Forschungszentrum J\"{u}lich COSY-FFE, the Georgian National Science
Foundation, and the CSC programme \#2011491103.

%
%


\begin{thebibliography}{00}
%
\bibitem{BUG1987} D.~V.~Bugg and C.~Wilkin, Nucl.\ Phys.\ A
    \textbf{467} (1987) 575.
%
\bibitem{ELL1987} C.~Ellegaard et al., Phys.\ Rev.\ Lett.\
    \textbf{59} (1987) 974.
%
\bibitem{KOX1993} S.~Kox et al., Nucl.\ Phys.\ A \textbf{556}
    (1993) 621.
%
\bibitem{CHI2006} D.~Chiladze et al., Phys.\ Lett.\ B
    \textbf{637} (2006) 170.
%
\bibitem{MCH2013} D.~Mchedlishvili et al., Eur.\ Phys.\ J.\ A
    \textbf{49} (2013) 49.
%
\bibitem{CAR1991} J.~Carbonell, M.~B.~Barbaro, and C.~Wilkin,
    Nucl.\ Phys.\ A \textbf{529} (1991) 653.
%
\bibitem{ARN2007} R.A.~Arndt, W.J.~Briscoe, I.I.~Strakovsky,
    R.L.~Workman, Phys.\ Rev.\ C \textbf{76} (2007) 025209;
    \verb=http://gwdac.phys.gwu.edu=.
%
\bibitem{KAC2005} A.~Kacharava, F.~Rathmann, C.~Wilkin,
    \emph{Spin Physics from COSY to FAIR}, COSY proposal
    \#\textbf{152} (2005), arXiv:nucl-ex/0511028.
%
\bibitem{MCH2013a} D.~Mchedlishvili, S.~Barsov, and
    C.~Wilkin, COSY proposal \#\textbf{218} (2013), available from\\
    \verb=http//www.collaborations.fz-juelich.de/ikp/anke=.
%
\bibitem{BAR2001} S.~Barsov et al., Nucl.\ Instrum.\ Methods A
    \textbf{462} (2001) 364.
%
\bibitem{MAI1997} R.~Maier et al., Nucl.\ Instrum.\ Methods A
    \textbf{390} (1997) 1.
%
\bibitem{MIK2013} M.~Mikirtychyants et al., Nucl.\
    Instrum.\ Methods A \textbf{721} (2013) 83.
%
\bibitem{GRI2007} K.~Grigoryev \emph{et al.}, AIP Conf.\ Proc.\
    \textbf{915} (2007) 979;\\ K.~Grigoryev \emph{et al.},
     Nucl.\ Instrum.\ Methods A \textbf{599} (2009) 130.
%
\bibitem{GRI2014} K.~Grigoryev et al., (\textit{in preparation}).
%
\bibitem{ENG2003} R.~Engels et al., Rev.\ Sc.\ Instrum.\
    \textbf{74} (2003) 4607.
%
\bibitem{SCH2003} R.~Schleichert et al., IEEE Trans.\
    Nucl.\ Sci.\ \textbf{50} (2003) 301.
%
\bibitem{HAJ1987} M.~Haji-Said et al., Phys.\ Rev.\ C
    \textbf{36} (1987)  2010.
%
\bibitem{ARV1984} J.~Arvieux et al., Nucl.\ Phys.\ A
    \textbf{431} (1984) 613.
%
\bibitem{ARV1988} J.~Arvieux et al., Nucl.\ Instrum.\ Methods
    A \textbf{273} (1988) 48.
%
\bibitem{CHI2006a} D.~Chiladze et al., Phys.\ Rev.\ ST Accel.\
    Beams \textbf{9} (2006) 050101.
%
\end{thebibliography}
\end{document}